\documentclass[letterpaper]{article} 
\usepackage{aaai2026}  
\usepackage{times}  
\usepackage{helvet}  
\usepackage{courier}  
\usepackage[hyphens]{url}  
\usepackage{graphicx} 
\usepackage{natbib}  
\usepackage{caption} 
\usepackage{algorithm}
\usepackage{algorithmic}

\usepackage{newfloat}
\usepackage{listings}
\DeclareCaptionStyle{ruled}{labelfont=normalfont,labelsep=colon,strut=off} 
\floatstyle{ruled}
\newfloat{listing}{tb}{lst}{}
\floatname{listing}{Listing}

\usepackage[switch]{lineno}
\usepackage{multirow}
\usepackage{amsmath,amssymb}
\usepackage{booktabs}
\usepackage{comment}
\usepackage{array}
\usepackage{framed}
\usepackage{textcomp}
\usepackage{xcolor}
\usepackage{tcolorbox}
\usepackage{pifont}
\usepackage{subcaption}
\usepackage{xspace}

\makeatletter
\DeclareRobustCommand\onedot{\futurelet\@let@token\@onedot}
\def\@onedot{\ifx\@let@token.\else.\null\fi\xspace}

\def\eg{\emph{e.g}\onedot} \def\Eg{\emph{e.g}\onedot}
\def\ie{\emph{i.e}\onedot} 
 
\def\etc{\emph{etc}\onedot}

\DeclareRobustCommand{\ourapproach}{{\sc PROD}\xspace}
\DeclareRobustCommand{\ourapproachbf}{{\sc \textbf{PROD}}\xspace}

\makeatother

\renewcommand{\paragraph}[1]{\vskip 0.01in \noindent {\bf #1.}}

\definecolor{SoftBlue}{RGB}{135, 206, 250}
\definecolor{SoftOrange}{RGB}{255, 224, 178}
\definecolor{SoftGreen}{RGB}{144, 238, 144}
\definecolor{CorrectGreen}{RGB}{76, 175, 80}
\definecolor{ErrorRed}{RGB}{211, 47, 47}

\title{Large Language Model Unlearning for Source Code}
\author{
Xue Jiang\textsuperscript{\rm 1,\rm 2}\thanks{Work done during Xue Jiang and Yihong Dong's internship at Tongyi Lab.}, 
Yihong Dong\textsuperscript{\rm 1,\rm 2,*}, 
Huangzhao Zhang\textsuperscript{\rm 3},
Tangxinyu Wang\textsuperscript{\rm 1},
Zheng Fang\textsuperscript{\rm 1}, \\
Yingwei Ma\textsuperscript{\rm 2}, 
Rongyu Cao\textsuperscript{\rm 2}, 
Binhua Li\textsuperscript{\rm 2}, 
Zhi Jin\textsuperscript{\rm 1}, 
Wenpin Jiao\textsuperscript{\rm 1}, 
Yongbin Li\textsuperscript{\rm 2}, 
Ge Li\textsuperscript{\rm 1}\thanks{Corresponding author.}}

\affiliations {
    \textsuperscript{\rm 1}Key Laboratory of High Confidence Software Technologies (Peking University), Ministry of Education; \\School of Computer Science, Peking University, Beijing, China \textsuperscript{\rm 2}Tongyi Lab, Alibaba Group \textsuperscript{\rm 3}Verdent AI\\
    \{jiangxue, dongyh\}@stu.pku.edu.cn, lige@pku.edu.cn
}

\begin{document}

\maketitle

\begin{abstract}
While Large Language Models (LLMs) excel at code generation, their inherent tendency toward verbatim memorization of training data introduces critical risks like copyright infringement, insecure emission, and deprecated API utilization, \etc. A straightforward yet promising defense is unlearning, \ie, erasing or down-weighting the offending snippets through post-training. However, we find its application to source code often tends to spill over, damaging the basic knowledge of programming languages learned by the LLM and degrading the overall capability. To ease this challenge, we propose \ourapproach for precise source code unlearning. \ourapproach surgically zeroes out the prediction probability of the prohibited tokens, and renormalizes the remaining distribution so that the generated code stays correct. By excising only the targeted snippets, \ourapproach achieves precise forgetting without much degradation of the LLM's overall capability. To facilitate in-depth evaluation against \ourapproach, we establish an unlearning benchmark consisting of three downstream tasks (\ie, unlearning of copyrighted code, insecure code, and deprecated APIs), and introduce Pareto Dominance Ratio (PDR) metric, which indicates both the forget quality and the LLM utility. Our comprehensive evaluation demonstrates that \ourapproach achieves superior overall performance between forget quality and model utility compared to existing unlearning approaches across three downstream tasks, while consistently exhibiting improvements when applied to LLMs of varying series. \ourapproach also exhibits superior robustness against adversarial attacks without generating or exposing the data to be forgotten. These results underscore that our approach not only successfully extends the application boundary of unlearning techniques to source code, but also holds significant implications for advancing reliable code generation.
\end{abstract}

\section{Introduction} 

Large Language Models (LLMs) have revolutionized the field of software engineering, automatically converting requirements into executable code with efficiency exceeding human programmers \cite{alphacode,Copilot,Agent4code,anthropic_claude_code,jiang2025coderl+}.
Such a leap is powered by the scaling law, \ie, ever-larger models and ever-growing corpora yield ever-better performance \cite{kaplan2020scaling,hoffmann2022training}.
However, the training corpora themselves are usually scraped from the open web, almost inevitably containing copyrighted fragments, vulnerable snippets, and deprecated Application Programming Interfaces (APIs), \etc
Because LLMs are trained autoregressively, they can memorize and later generate these undesirable snippets \cite{dong-etal-2024-generalization}, exposing users to legal risk, security breaches, fragile legacy code, and so on.
Empirical studies show that up to 40\% of LLM-generated code contains exploitable vulnerabilities \cite{safecoder}, while a wave of high-profile lawsuits accuses leading LLMs of violating the copyright of the developers \cite{zhang2024right,butterick2022github}.

Purging every problematic code snippet from training corpora and retraining the LLMs from scratch would be straightforward and ideal, yet the cost is obviously unacceptable.
A practical alternative is to force the LLMs to selectively forget what it was never meant to know, which is the objective known as unlearning \cite{Unlearning,DBLP:journals/natmi/LiuYJCBHYLXLVBKL25}.
Given a functional LLM, unlearning algorithms behave like erasers:
\ding{182} they excise the influence of the undesirable samples from the target LLM,
while \ding{183} leaving all other knowledge and capabilities intact.
Existing LLM unlearning methods \cite{Unlearning,NPO,DPO,FLAT} typically optimize the model to suppress the likelihood of undesirable outputs by reversing the gradient descent process upon undesirable contents or setting them as negative samples \cite{DBLP:journals/natmi/LiuYJCBHYLXLVBKL25,Unlearning,NPO,DPO,FLAT}.
\Eg, Gradient Ascent (GA) \cite{Unlearning} tunes parameters uphill on samples to be forgotten.
For natural language tasks, such methods have been demonstrated effective in excising toxic responses, secret leakage, or policy-violating answers without impairing general capabilities significantly.

Programming languages, however, are bound by rigid syntax, along with strict semantics.
When existing unlearning methods directly down-weighting undesirable code snippets, they also strip the language scaffolding that keeps the surrounding code valid.
It results in direct utility loss: syntax errors and grammar errors occur, high-frequency tokens repeat, or even the LLMs fall mute.
Our pilot empirical study confirms such an issue (see Figure \ref{fig:case}): after manipulated by existing unlearning techniques, the target LLMs frequently violate programming language rules or emit nonsensical code snippets.
We argue that the root cause here is granularity, \ie, current techniques erase entire slices of knowledge (\ie, the undesirable snippets), damaging the foundational grammar and universal coding patterns the model once mastered through pre-training.
Therefore, the open challenge is to eliminate only the offending snippets with surgical precision (maximize forget quality, \ie, the degree that the undesirable contents are eliminated) while leaving model utility (\ie, the general performance of the target LLM) intact.

We introduce \textbf{P}robabilistic \textbf{R}edistribution for \textbf{O}utput \textbf{D}istribution (abbreviated as \ourapproach), a surgical, code-oriented unlearning method that forgets code snippets that must be forgotten while leaving all other knowledge of programming languages intact.
Unlike existing unlearning approaches, which operate at the granularity of samples, \ourapproach manipulates distributions over individual tokens in a fine-grained granularity.
\ourapproach first captures the full distribution over vocabulary predicted by the target LLM, then surgically zeroes out every probability mass assigned to tokens in the undesirable snippet.
After pruning incidental noises, it reallocates the probability distribution across the remaining vocabulary in a manner that faithfully preserves statistics of programming languages learned by the target LLM. By optimizing the model to match this carefully sculpted target distribution, \ourapproach attains near-perfect forgetting of the targeted code while effectively maintaining its utility unchanged. We also propose a benchmark for source code unlearning along with a metric, Pareto Dominance Ratio (PDR).
The benchmark covers three tasks:
\ding{182} copyrighted code unlearning, where the goal is to remove copyrighted code snippets from LLMs;
\ding{183} insecure code unlearning, which aims to avoid security vulnerabilities produced by LLMs;
\ding{184} deprecated API unlearning, which prevents LLMs from generating APIs from outdated libraries or packages.
PDR metric evaluates unlearning approaches by considering both forget quality and model utility.

Our in-depth evaluation over the three unlearning tasks shows that \ourapproach achieves the best PDR score, suggesting its best overall performance between forget quality and model utility.
Experiments on four distinct LLMs (\ie, CodeLlama-7B, Qwen2.5-Coder-7B, Deepseek-coder-6.7, Starcoder-7B) demonstrate the broad applicability of \ourapproach.
Adversarial attack experiment further reveals that \ourapproach is the only unlearning method among existing LLM unlearning methods resilient against targeted attacks without reproducing forgotten codes.

Our main contributions are highlighted as follows:
\begin{itemize}

    \item We conduct an investigation of existing LLM unlearning approaches on code-related tasks, revealing that they lead to severe model utility degradation, making models practically unusable for code generation.

    \item We propose a novel unlearning approach \ourapproach, that can precisely forget undesired code snippets and preserve knowledge of programming languages intact in LLMs.
    
    \item We identify three critical unlearning applications in the code generation domain (including copyrighted code unlearning, insecure code unlearning, and deprecated API unlearning) and establish a benchmark for evaluating LLM unlearning approaches for code.
    
    \item Extensive experimental results demonstrate that \ourapproach significantly outperforms existing LLM unlearning approaches, achieving successful forgetting of specific code while effectively preserving code generation capabilities. Our source code and data are available at \url{https://github.com/jiangxxxue/PROD}.
    
\end{itemize}

\section{Related Work}
In this section, we outline the most relevant work of \ourapproach and detail the extended related work section in Appendix.

\subsection{Code Generation with LLMs}
Code generation has been significantly advanced by LLMs, from general-purpose LLMs like GPT-4 \cite{achiam2023gpt4} to a variety of specialized code LLMs \cite{codex, CodeLlama, zhu2024deepseekcoder, hui2024qwen2.5-coder, team2024codegemma}.
However, their training on vast, public code repositories introduces critical challenges regarding security, reliability, and legality.
A primary concern is that LLMs may replicate vulnerabilities. To mitigate this, researchers have explored security-focused fine-tuning with curated datasets (SaferCode \cite{safecoder}) and in-context learning with secure examples \cite{mohsin2024can}. Beyond security, model reliability is undermined by poor handling of software versions. Studies show that LLMs often suggest deprecated APIs (SecuCoGen \cite{wang2024and}) and struggle with API changes across different library releases, a challenge highlighted by the VersiCode benchmark \cite{wu2024versicode}. Finally, legal issues arise from models generating code that may violate software licenses, as systematically evaluated by LiCoEval \cite{xu2024first}.

These works highlight that despite remarkable progress, LLM-based code generation still falls short of developers' expectations for producing secure, legally compliant, and up-to-date code. Our work aims to address this gap by proposing a general unlearning approach for suppressing undesired code output from LLMs.

\subsection{LLM Unlearning}
Machine unlearning aims to remove the influence of specific data from a trained model \cite{DBLP:conf/sp/CaoY15}. Initial research primarily focused on classification models~\cite{DBLP:conf/cvpr/GolatkarAS20,DBLP:conf/aistats/IzzoSCZ21,DBLP:conf/sp/BourtouleCCJTZL21}, with methods broadly categorized into data-reversed training~\cite{DBLP:journals/tnn/TarunCMK24,DBLP:journals/tifs/ChundawatTMK23}, influence function-based approaches~\cite{DBLP:conf/aistats/IzzoSCZ21}, and optimization-based unlearning~\cite{DBLP:conf/icml/GuoGHM20,DBLP:conf/alt/Neel0S21}.

The vast scale and generative nature of LLMs present unique challenges, rendering many traditional methods impractical~\cite{DBLP:journals/natmi/LiuYJCBHYLXLVBKL25}. Specifically, the computational cost of Hessian inversion makes influence function-based methods prohibitive for LLMs. Consequently, recent work has shifted towards optimization-based techniques tailored for LLMs in the NLP domain. Methods such as GA~\cite{Unlearning}, DPO-based unlearning~\cite{DPO}, NPO~\cite{NPO}, and FLAT~\cite{FLAT} have been proposed to erase private data, copyrighted material, and harmful content. The evaluation of these techniques relies on specialized benchmarks~\cite{DBLP:journals/corr/abs-2401-06121, DBLP:journals/corr/abs-2310-10683} and is often tested for robustness against adversarial attacks like prefix injections~\cite{Jailbroken,DBLP:journals/corr/abs-2408-10682,qi2025safety}.

Given that source code possesses unique characteristics compared to natural language, although the exploration of LLM unlearning in NLP has proven its value, its application in code generation remains largely unexplored.

\section{Preliminary of LLM Unlearning}

In this section, we first formalize the LLM unlearning problem and establish the evaluation criteria.
Next, we briefly survey four representative baselines that span current LLM unlearning paradigms.
Finally, we conduct an empirical pilot study revealing that, on source code, existing baselines cause great losses in model utility, motivating the need for a more surgical solution.

\paragraph{Objective of Unlearning}
Within the broad landscape of machine unlearning, we focus on optimization-based methods.
Here, unlearning is a post-training procedure that removes the LLM's ability to produce specific, undesirable code snippets.
Let $\mathcal{D}_f=\{x_f^{(i)},y_f^{(i)}\}_{i=1}^{N_f}$, denote the set of such snippets, where each pair consists of a prompt $x_f^{(i)}$, and its undesirable completion $y_f^{(i)}$ (note that the prompt $x_f^{(i)}$ can be empty in cases such as the entire snippet $y_f^{(i)}$ must be forgotten).
After unlearning, the target LLM $\pi_\Theta$ is supposed to assign zero probability to every undesirable $y_f^{(i)}$ given $x_f^{(i)}$.
Formally, we quantify this objective with the loss $\mathcal{L}_f\left(\pi_\Theta(x_f),y_f\right)$ that measures the divergence between the LLM's output $\pi_\Theta(x_f^{(i)})$ and the undesirable output $y_f^{(i)}$.
Unlearning is then cast as the following optimization:

\begin{equation}
    \min_{\Theta}\sum_{i=1}^{N_f}-\mathcal{L}_f\left(\pi_\Theta(x_f^{(i)}), y_f^{(i)}\right), \label{eq:unlearning_formulation}
\end{equation}
where maximizing $\mathcal{L}_f$ forces $\pi_\Theta$ to forget every undesirable $y_f^{(i)}$ in $\mathcal{D}_f$, including copyrighted code, insecure fragment, or deprecated API invocation, \etc
Notably, $\mathcal{D}_f$ merely captures behaviors we wish to eliminate, and samples in it do not have to appear verbatim in the original training data.

\paragraph{Metrics}
As aforementioned, unlearning requires to erase only what is undesirable or even harmful, and retain everything else.
We therefore assess unlearning approaches along two axes:
\ding{182} Forget quality: the thoroughness with which $\pi_\Theta$ eliminates the influence of $\mathcal{D}_f$, measured by the degree of dissimilarity between the output of $\pi_\Theta$ and $y_f^{(i)}$ when prompted with $x_f^{(i)}$.
\ding{183} Model utility: the collateral side damage caused by the procedure, gauged by $\pi_\Theta$’s retained performance on other general tasks and datasets.
Assessing the balance between forget quality and model utility is essential due to their trade-off in unlearning methods.

\begin{figure}[t!]
    \centering
    \includegraphics[width=1\linewidth]{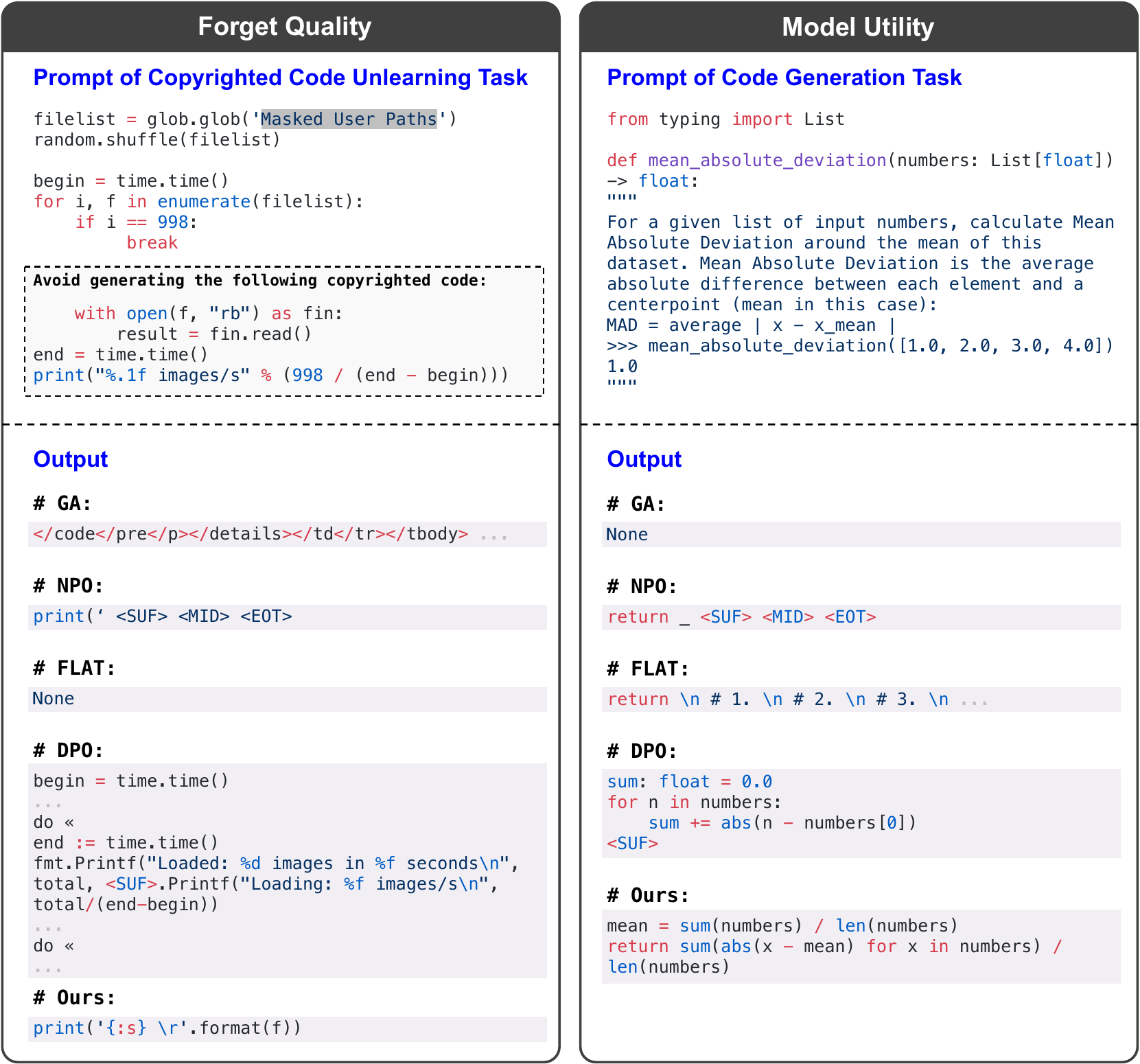}
    \caption{A case on the forget quality and model utility of existing method in unlearning code. Existing methods (GA, NPO, FLAT, DPO) exhibit severe utility degradation, such as mute refusal, token collapse, or syntactic incoherence. \ourapproach successfully unlearns copyrighted content while maintaining utility for general code generation tasks.}
    \label{fig:case}
\end{figure}

\paragraph{LLM Unlearning Solutions}
Existing methods are almost exclusively tuned to optimize the forget objective in Equation (\ref{eq:unlearning_formulation}), and treat downstream utility as an after-thought, resulting in plausible damage.
\ding{182} Gradient Ascent (GA) \cite{Unlearning} carries out a ``reverse'' of training by ascending the loss on $\mathcal{D}_f$, pushing $\pi_\Theta$ away from the parameter region that supports undesirable generations.
\ding{183} Negative Preference Optimization (NPO) \cite{NPO}, which is a variant of Direct Preference Optimization (DPO) \cite{DPO}, supplies only negative examples (\ie, $(x_f^{(i)}, y_f^{(i)})$) to repel $\pi_\Theta$ from the unwanted distribution.
\ding{184} DPO is a general preference optimization method that, while not specifically designed for unlearning, can be adapted for this purpose by treating the content-to-be-forgotten as the ``dispreferred" response.
\ding{185} Forget Data Only Loss Adjustment (FLAT) \cite{FLAT} formulates $\mathcal{L}_f$ with $f$-divergence, enforcing $\pi_\Theta$ to assign maximal probability to a hand-crafted refusal template (\eg, ``I do not know'') while minimizing probability on each $(x_f^{(i)}, y_f^{(i)})$.

\begin{figure*}[t!]
    \centering
    \includegraphics[width=0.98\textwidth]{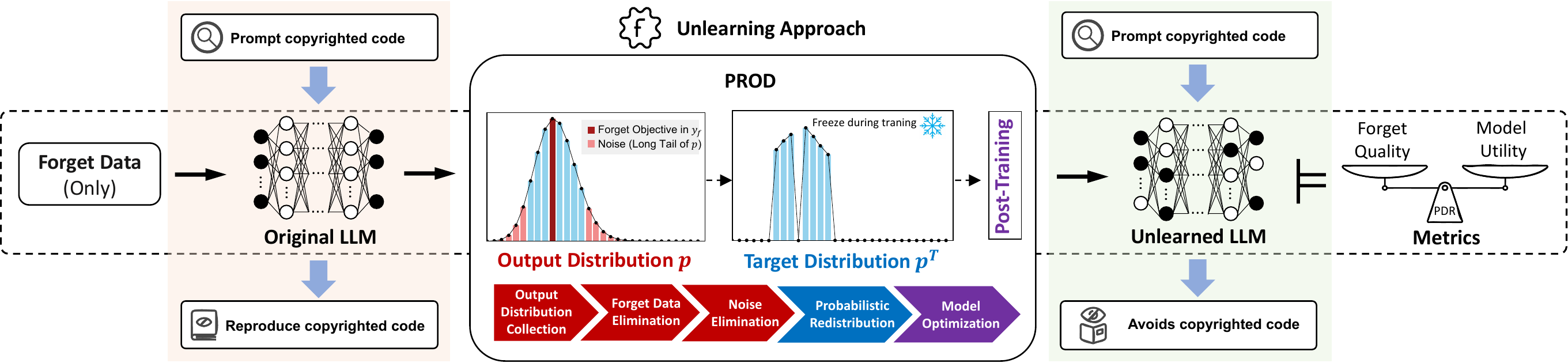}
    \caption{The \ourapproach unlearning pipeline.
    The pipeline contains three key steps: \ding{182} Suppress the probabilities of undesirable code snippets to zero. \ding{183} Redistribute the probabilities across the remaining vocabulary. \ding{184} Optimize the model to match the surgically sculpted target distribution.}
    \label{fig:workflow}
\end{figure*}

\paragraph{A Motivating Example}
To vividly show the collateral damage caused by existing unlearning methods on source code tasks, we conduct an empirical study.
We benchmark the above-mentioned baselines on a copyrighted code unlearning task and then measure their retained utility with a standard code generation benchmark (detailed setups in the experiment section).
As Figure~\ref{fig:trade_off} indicates, all baseline approaches achieve $\approx90\%$ forgetting ratio, yet their success ratio on code generation degrades to near-zero when they reach best forget quality.
In other words, these approaches successfully erase the undesirable snippets (\ie, copyrighted code) but simultaneously cause LLMs to forget how to program, \ie, such an outcome makes the target LLMs almost useless.
Our motivation in this paper is precisely the opposite, as we argue to excise the undesirable code snippets while leaving the target LLMs fully operational.

A closer look at the baselines’ outputs (a case is presented in Figure\ref{fig:case}) reveals three dominant failure modes:
\ding{182} Mute Refusal: the model declines to produce code,
\ding{183} Token Collapse: the output degenerates into repetitive high-frequency tokens such as ``\textbackslash n'' or training artifacts like ``\texttt{<}SUF\texttt{>}'',
and \ding{184} Syntactic Incoherence: the generated code violates fundamental language rules and fails to compile or execute.
These breakdowns arise from the baselines’ coarse granularity.
By bluntly erasing every trace of $y_f^{(i)}$ from $\pi_\Theta$, they eliminate not only the undesirable snippet but also the surrounding “scaffolding”.
Natural languages tolerate such collateral damage as they are much more flexible, yet programming languages, which are governed by rigid syntax and precise dependencies, do not.
Once the supporting knowledge and structure are disturbed, the entire code generation capability collapses.
To preserve utility while still forgetting, we introduce \ourapproach, a surgical unlearning solution that manipulates the output distribution at token-level granularity instead of bluntly excising entire snippet sequences.

\section{\ourapproachbf}
To protect model utility while guaranteeing forgetting, we introduce \ourapproach, a surgically precise unlearning method that manipulates the LLM’s token-level output distributions.
As illustrated in Figure \ref{fig:workflow} and formalized in Algorithm \ref{alg:pipeline}, the working pipeline of \ourapproach consists of three steps.
\ding{182} Suppress the probabilities predicted by the target LLM of tokens that constitute the undesirable snippet to zero (\ie, remove them from the output).
\ding{183} Redistribute the probabilities across the remaining vocabulary so that the resulting distribution mirrors the original statistical patterns and preserves knowledge that the LLM learn from the code corpora.
\ding{184} Optimize the target LLM to match this surgically sculpted target distribution.
As a result, \ourapproach achieves complete erasure of undesirable or even harmful code with virtually no collateral damage to any other programming competency.

\subsection{Target Distribution Sculpting}
We begin by sculpting the target LLM’s output distribution into a precise supervisory signal that will steer the entire unlearning process of \ourapproach.

\paragraph{Output Distribution Collection}
The initial step of \ourapproach retrieves the LLM’s raw next-token distributions for each $(x_f, y_f)$ pair \footnote{For simplicity, the superscripts in $x_f^{(i)}$ and $y_f^{(i)}$ are omitted henceforth in this section.}.
Given a forget pair $(x_f,y_f)$ with $y_f=(y_{f,1},\cdots,y_{f,L})$, at time-step $t$, we feed the target LLM ($\pi_\Theta$) the concatenation of the prompt $x_f$ and the preceding ground-truth prefix $y_{f,<t}=(y_{f,1},\cdots,y_{f,t-1})$ to obtain the logits $h_t\in\mathbb{R}^{|\mathcal{V}|}$, where $\mathcal{V}$ refers to the vocabulary.
A softmax converts the logits into the original distribution $p_t(\cdot|x_f,y_{f,<t})$. 
$p_t$ then serves as the canvas where the subsequent manipulations will be carried out.

\paragraph{Forget Data Elimination}
To prohibit any generation of the target sequence, \ourapproach performs a single, decisive edit on $h_t$ at every $t$.
The logit entry corresponding to the target token $y_{f,t}$ is suppressed to $-\infty$ as below:

\begin{equation}
\hat{h}_t[j] = \begin{cases}
-\infty, & \text{If~}\mathcal{V}[j]=y_{f,t}, \\
h_t[j], & \text{Otherwise},
\end{cases} \label{eq:logit_suppression_revised}
\end{equation}
where $[\cdot]$ denotes vector indexing, and $\mathcal{V}[j]$ refers to the $j$-th token in the whole vocabulary $\mathcal{V}$.
A softmax normalization of $\hat h_t$ would yield the eliminated distribution $\hat{p}_t(\cdot|x_f,y_{f,<t})$, which assigns zero probability to $y_{f,t}$ and redistributes its entire mass across the remaining vocabulary in exact proportion to their original likelihoods.

\paragraph{Noise Elimination}
Suppressing the target token $y_{f,t}$ can inadvertently amplify less probable, non-target tokens,  injecting noise into $\hat{p}_t(\cdot|x_f,y_{f,<t})$.
We therefore prune the tail (\ie, the noise) via nucleus sampling \cite{Holtzman} with the threshold of $p$.
Let $\mathcal{S}_p$ denote the smallest set of tokens whose cumulative probability under $\hat p_t$ reaches $p$.
All logits outside $\mathcal{S}_p$ are then set to $-\infty$:

\begin{equation}
\widetilde{h}_t[j]= 
\begin{cases} 
\hat{h}_t[j], & \text{If~}\mathcal{V}[j]\in\mathcal{S}_p, \\
-\infty, & \text{Otherwise.}
\end{cases}
\label{Eq: Noise Elimination}
\end{equation}
Only tokens within $\mathcal{S}_p$ receive redistributed probability, and the rest are eliminated as noise. The softmax normalization upon $\widetilde{h}_t$ results in noise-trimmed distribution $\widetilde{p}_t$.

Crucially, forget data elimination must precede noise elimination.
Reversing the order would either retain target tokens or produce an overly sparse distribution. This two-step sequence ensures a clean unlearning process while preserving a robust, information-rich distribution over the surviving vocabulary.

\paragraph{Probabilistic Redistribution}
After excising both the target token $y_{f,t}$ and the noisy tail outside $S_p$, $\widetilde{h}_t$ holds the remaining probability mass. \ourapproach redistributes this mass across the surviving safe vocabulary (\ie, $\mathcal{S}_p$) through softmax, and therefore adopts $\widetilde{p}_t$ as the supervisory signal $p_T$.

\subsection{Post-Training}
With the sculpted supervisory distribution $p_T=\widetilde{p}_t$, \ourapproach optimizes the target LLM $\pi_\Theta$ via standard cross-entropy loss, nudging its outputs to align with this sculpted distribution while preserving every other learned competency.

\paragraph{Objective}
The optimization objective of \ourapproach is to minimize the loss $\mathcal{L}_{\text{\ourapproach}}$, finalized as following:

\begin{equation}
\mathcal{L}_{\text{\ourapproach}}(x_f,y_f;\pi_\Theta) = -\sum_{t=1}^L\sum_{w\in\mathcal{V}}p_T(w|\cdot)\log \pi_\Theta(w|\cdot), \label{eq:loss_function}
\end{equation}

\noindent where $\pi_\Theta(\cdot)$ refers to the actual output distribution of $\pi_\Theta$. The conditions (\ie, $x_f,y_{f,<t}$) are neglected for simplicity. In default design, \ourapproach adopts cross-entropy. The framework is agnostic, \ie, KL \cite{kullback1951}, JS \cite{lin1991}, or any other distributional divergence can be substituted seamlessly.

\paragraph{Optimization}
Any compatible optimizer, with gradient descent included, can be employed in \ourapproach.

\begin{algorithm}[t]
\caption{Pseudocode for \textbf{\ourapproach}.}
\label{alg:pipeline}
\small{
\begin{algorithmic}[1]
\REQUIRE LLM $\pi_\Theta$; forget set $\mathcal{D}_f$; learning rate $\eta$; training epochs $N$; hyperparameters $p, \alpha$.
\ENSURE Unlearned LLM $\pi_\Theta^*$.
\FOR{epoch k $= 1$ $\textbf{to}$ $N$}
    \FOR{each $(x_f, y_f) \in \mathcal{D}_f$}
        \STATE Collect output distribution $p(w|x_f, y_{f,<t})$.
        \STATE Compute $\hat{h}_t$ to apply forget data elimination.
        \STATE Compute $\widetilde{h}_t$ to perform noise elimination.
        \STATE Compute probabilistic redistribution $p_\text{T}$.
        \STATE Calculate loss function $\mathcal{L}_{\text{PROD}}$.
        \STATE $\Theta_{k+1} = \Theta_k - \eta \nabla_{\Theta} \mathcal{L}_{\text{PROD}}.$
        \STATE $k \leftarrow k+1$ and $\Theta^* \leftarrow \Theta_{k}$
    \ENDFOR
\ENDFOR
\RETURN $\pi_\Theta^*$
\end{algorithmic}}
\end{algorithm}

\subsection{Key Know-How of \ourapproachbf}

In general, \ourapproach provides two strengths.
\ding{182} Model utility preservation: the supervisory $p_T$ is a surgically sculpted clone of the original $p_t$, excising only the undesirable or forbidden snippets while leaving every other competency untouched.
\ding{183} Efficient convergence: since $p_T$ remains close to the $\pi_\Theta$'s initial output, the optimization process is fast, stable, and light on compute. Furthermore, we elaborate on two implementation details.

\paragraph{$\mathbf{\alpha}$-suppression} 
Although PROD excels in unlearning long code, early experiments revealed that short forget sequences $y_f$ often require more training steps since the supervisory signal is weak.
To modulate the ``forget strength'', we introduce the $\alpha$-suppression trick and modify the supervisory signal $p_T$ as follows:

\begin{equation} \label{eq:target_distribution_revised}
   p_\text{T}(w|x_f, y_{f,<t})=
   \begin{cases}
        -\alpha \cdot p_o(w|\cdot), & \text{If~}w=y_{f,t}, \\
        \widetilde{p}_t(w|\cdot), & \text{Otherwise}, \\
    \end{cases}
\end{equation}
where $p_o$ and $\widetilde{p}_t$ refers to the original and noise-trimmed distributions, respectively. Still, some conditions are neglected for better presentation.
$\alpha$ is a hyperparameter to control the degree of forget, \ie, setting $\alpha=0$ disables the trick ($p_T=\widetilde{p}_t$); increasing $\alpha$ amplifies the negative pressure on $y_f$, yielding a stronger unlearning signal.
Note that while $\alpha$-suppression introduces negative probability, it is harmless, or even beneficial, under the setting of cross-entropy as defined in Equation \ref{eq:loss_function}:
\ding{182} the remaining terms maintain the loss magnitude and gradient direction;
and \ding{183} the negative entry amplifies the penalty on the forget token without computational issues.

\paragraph{Training Stability}
As training proceeds, $\pi_\Theta$’s distribution drifts with every gradient step.
If we recollected $p_o$ and recomputed $p_T$ on-the-fly, the supervisory signal would swing constantly, destabilizing the training process and risking collapse.
To avoid such an issue, we freeze $p_o$ at its initial state, \ie, the distributions produced by the target LLM before unlearning, thereby guaranteeing a steady and reliable target throughout optimization.

\begin{figure*}[ht!]
    \centering
    \begin{subfigure}[b]{0.32 \textwidth}
        \centering
        \includegraphics[width=\textwidth]{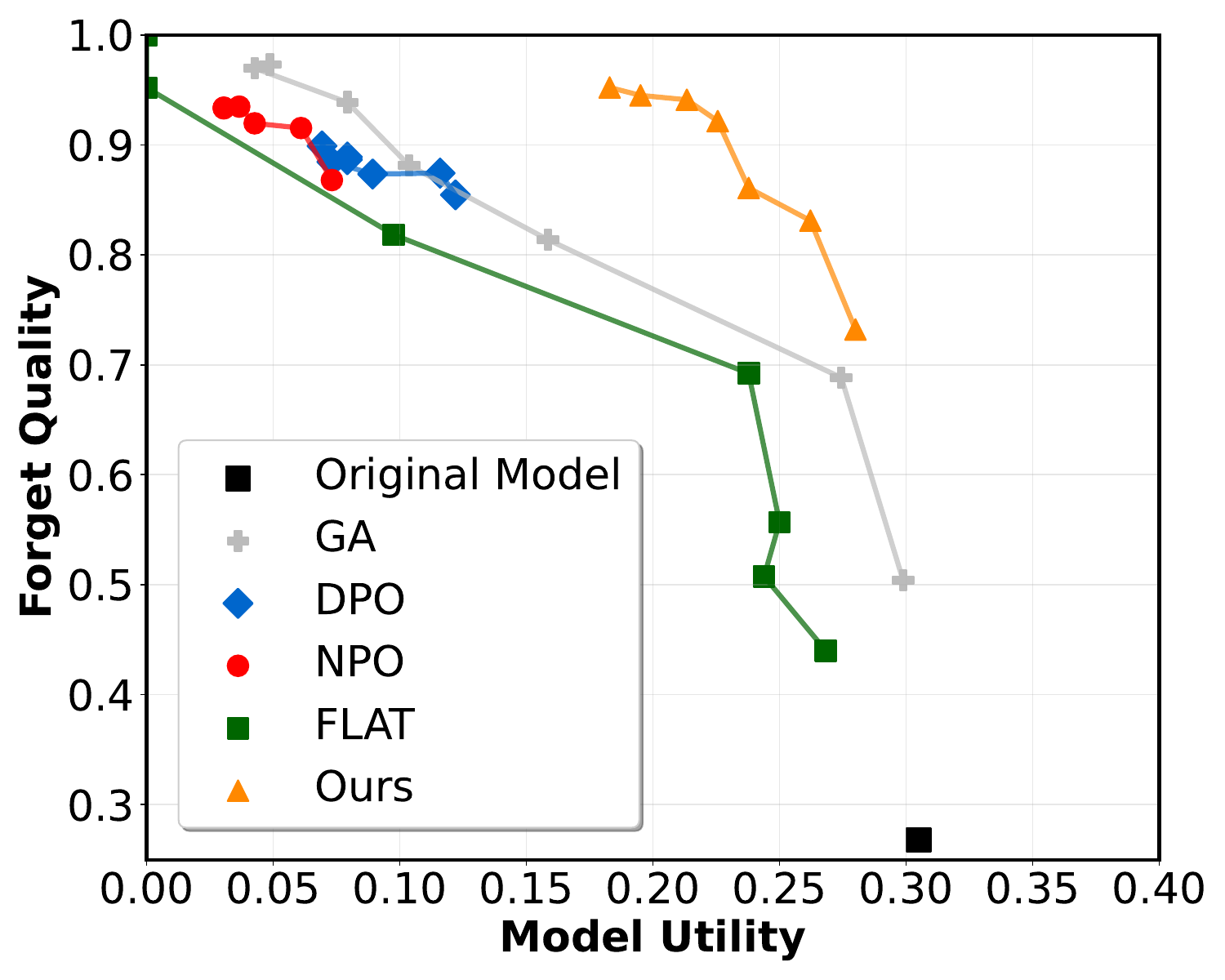}
        \caption{Copyrighted Code Unlearning}
        \label{copyright_trade_off}
    \end{subfigure}
    \hspace{0.005\textwidth}
    \begin{subfigure}[b]{0.32 \textwidth}
        \centering
        \includegraphics[width=\textwidth]{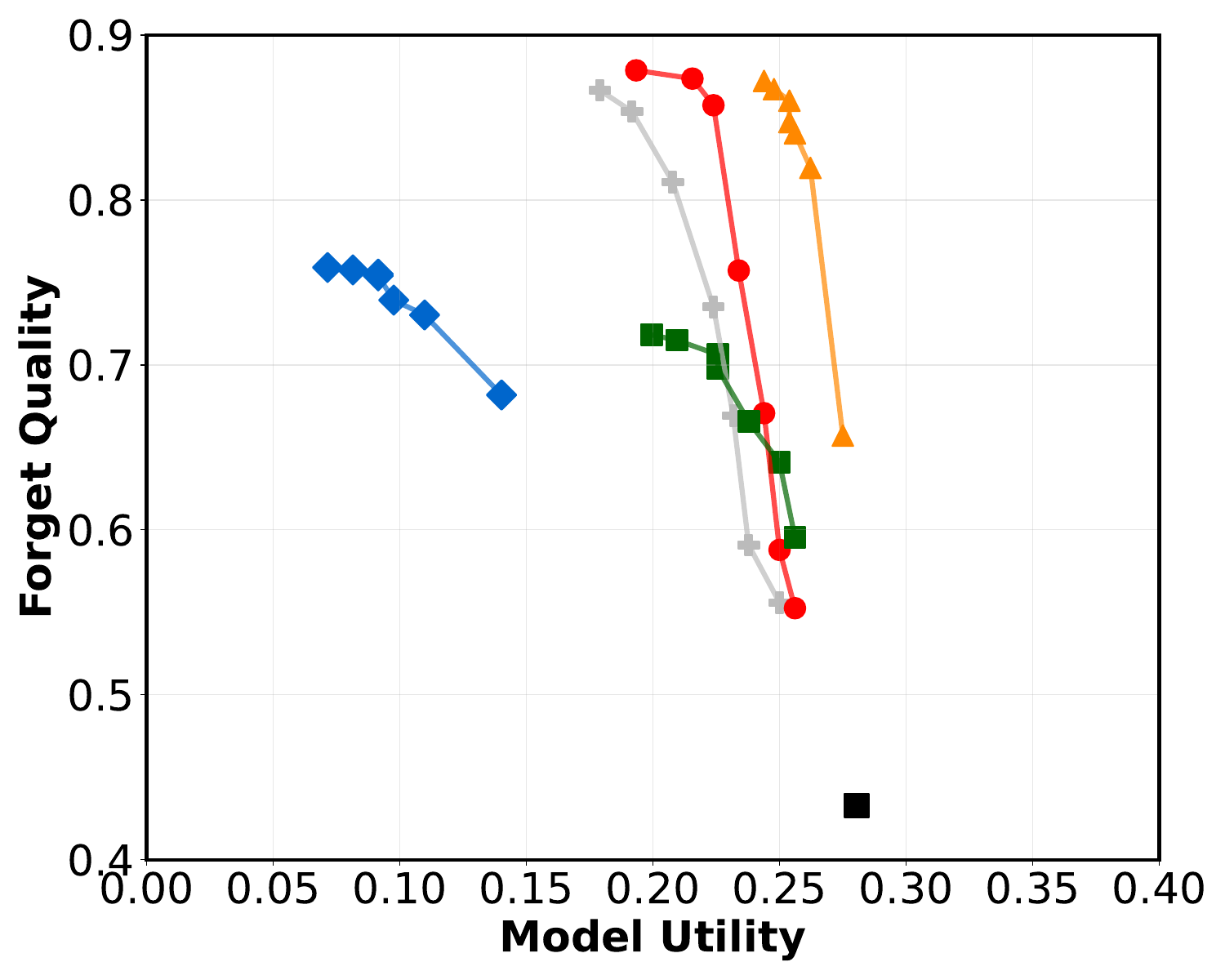}
        \caption{Insecure Code Unlearning}
        \label{fig:secure_trade_off}
    \end{subfigure}
    \hspace{0.005\textwidth}
    \begin{subfigure}[b]{0.32 \textwidth}
        \centering
        \includegraphics[width=\textwidth]{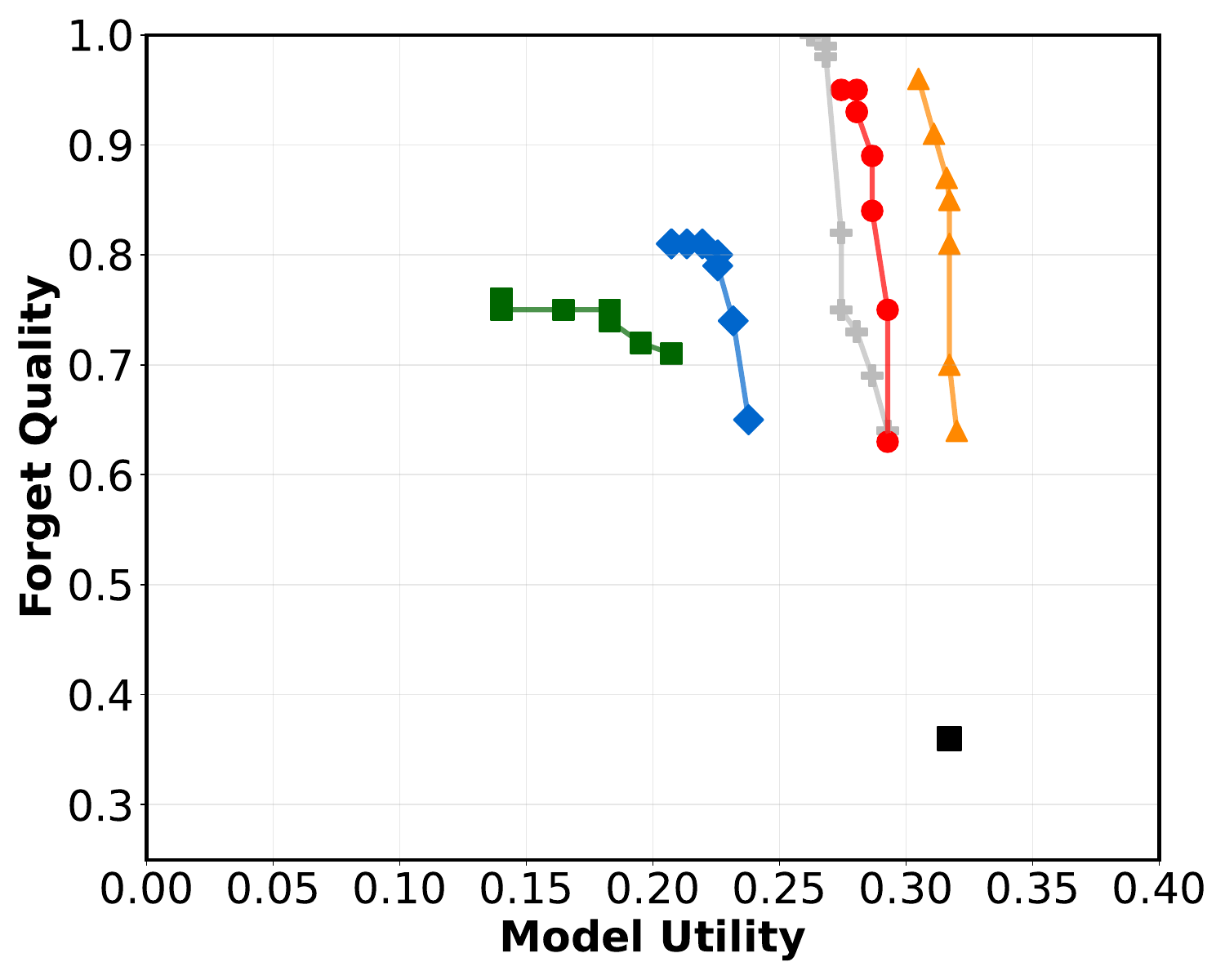}
        \caption{Deprecated API Unlearning}
        \label{fig:API_trade_off}
    \end{subfigure}
    \caption{Forget quality versus model utility across different unlearning tasks. Performance curves positioned closer to the upper-right corner indicate superior approach effectiveness.}
    \label{fig:trade_off}
\end{figure*}

\section{Evaluation}
We introduce a dedicated benchmark for code unlearning and report the main experimental results in this section.
Please refer to the appendix for full implementation details and a perceptual quality study.

\subsection{Code Unlearning Benchmark} \label{sec:benchmark}
Current unlearning benchmarks focus almost exclusively on natural language, leaving code generation unexamined.
To fill this gap, we curate and adapt datasets into a code unlearning benchmark, paired with custom metrics.
The benchmark contains three unlearning tasks (\ie, copyrighted code, insecure code, and deprecated APIs) to measure forget quality and a general code generation set to assess model utility.

\paragraph{Copyrighted Code Unlearning}
We randomly draw $100$ files from the high-quality, deduplicated Stack corpus \cite{TheStack} to serve as the forget set $\mathcal{D}_f$, simulating scenarios where users wish to purge copyrighted code files from an LLM.
To mimic copyright protection, test files, toy examples, and standard templates are filtered and discarded, retaining those with expressive form, such as substantive implementations, business logic, and non-trivial algorithms.
During unlearning, $x_f$ is empty and $y_f$ contains the full content of the retrieved file; when assessing forget quality, we prompt the target LLM with the first half of each file and evaluate against the other half.
As the law of copyright protects expression instead of ideas \cite{autry2002toward,copyright_protect}, textual similarity is employed as an indicator of potential copyright infringement.
Thus, the forget quality score is measured by the complement of BLEU \cite{BLEU} (\ie, $1-\text{BLEU}$), where higher values indicate stronger erasure of the protected code snippets.

\paragraph{Insecure Code Unlearning}
CyberSecEval from Purple Llama \cite{PurpleLlama} is a cybersecurity benchmark, containing 1,916 code snippets collected from open sources that exhibit 50 distinct CWE vulnerabilities \cite{cwe} across eight programming languages.
We adopt these snippets as our forget set $\mathcal{D}_f$.
Evaluation proceeds in an autocomplete setup, \ie, given the code that precedes a known vulnerable segment (\ie, $x_f$), we prompt the model to continue and then inspect the continuation for security flaws (\ie, $y_f$). 
Forget quality is assessed by the average of two complementary indicators:
\ding{182} $1-\text{BLEU}$ between the LLM's generation and the vulnerable snippet,
and \ding{183} the pass rate under CyberSecEval’s built-in static analyzer, which detects any insecure lines.

\paragraph{Deprecated API Unlearning}
This task is curated atop Versicode \cite{wu2024versicode}, which is a version-aware code-generation dataset.
With the original dataset, we apply three additional preprocessing steps.
\ding{182} Version filtration discards packages with ambiguous version constraints (\eg, \texttt{>=2.1.0}), retaining those with explicit version requirements (\eg, \texttt{==2.1.0}).
\ding{183} File filtration removes packages that have less than 3 snippets.
\ding{184} Temporal split designates a middle release as the deprecation boundary, labeling all invocations of APIs prior to that release as “deprecated” and the remainder as “still-valid.”
The final dataset consists of 252 packages, 3,449 snippets in total ($\approx$14 snippets and 10 distinct APIs per package).
Unlearning is carried out on the deprecated APIs, and evaluated on the remaining still-valid ones.
Given the code context immediately preceding an API call, we measure forget quality by exact-match accuracy.

\paragraph{General Code Generation}
Following prior work \cite{safecoder}, we estimate model utility through HumanEval \cite{codex}, which is a widely used code generation benchmark.
We employ functional correctness verified against test cases as the indicator of model utility.

\paragraph{Overall Performance Indicator}
Forget quality and model utility often pull in opposite directions; naively averaging them or even independently assessing them overlooks the real trade-off.
To make this multi-objective comparison explicit, we introduce the Pareto Dominance Ratio (PDR).
Let $\mathcal{M}=\{m_1,\cdots,m_n\}$ be the set of evaluated unlearning methods and  $\mathcal{O}=\{\text{forget, utility}\}$ the two objectives.
For any method $m_i\in\mathcal{M}$, PDR is defined as following:

\begin{equation}
\text{PDR}(m_i)=\frac{\bigl|{m_j\in\mathcal{M}\setminus{m_i}\mid m_i\succ_{\mathcal{O}} m_j}\bigr|}{|\mathcal{M}|-1},
\end{equation}

\noindent where $m_i\succ_\mathcal{O} m_j$ means that $m_i$ Pareto-dominates $m_j$, \ie, $m_i$ is no worse on both objectives and strictly better on at least one than $m_j$.
Geometrically, on a 2-D scatter of forget quality versus model utility,  
$m_i\succ_{\mathcal{O}}m_j$ places $m_i$ strictly to the upper-right of $m_j$.
PDR therefore reports the fraction of rival methods that $m_i$ dominates, yielding a single, interpretable summary of overall performance.

\begin{table}[t]\footnotesize
\centering
\caption{PDR (in $\%$) among unlearning approaches.}
\label{tab:unlearning_comparison}
\begin{tabular}{lccccc}
\toprule
\textbf{Task} & \textbf{GA} & \textbf{DPO} & \textbf{NPO} & \textbf{FLAT} & \textbf{Ours} \\
\midrule
Copyright & 15.3 & 3.6 & 0.6 & 0.0 & \textbf{41.8} \\
Insecurity & 13.3 & 0.0 & 24.5 & 6.1 & \textbf{70.4} \\
Deprecation & 31.5 & 20.6 & 51.6 & 0.0 & \textbf{58.0} \\
\bottomrule
\end{tabular}
\label{tab:PDR}
\end{table}

\subsection{Experimental Results}

\paragraph{Unlearning Performance}
We benchmark \ourapproach against four representative baselines (\ie, GA, DPO, NPO, and FLAT) on all three unlearning tasks.
The target model is CodeLlama-7B \cite{CodeLlama}.
Greedy decoding (zero temperature) is applied for every inference run. On each task, we compute PDR with forget quality (task-specific) and overall model utility (via HumanEval).
Comprehensive results are presented in Table \ref{tab:PDR} and illustrated in Figure \ref{fig:trade_off}.

\ourapproach consistently occupies the upper-right region of the scatter plot (Figure \ref{fig:trade_off}), simultaneously delivering stronger forget quality and higher model utility than the baselines.
The PDR metric (Table \ref{tab:PDR}) also quantifies this dominance -- \ourapproach tops every task, yielding an average relative gain of 124\% over the strongest competitor. Training dynamics offer further insight. By analyzing the training log, we find that \ourapproach forgets rapidly during early steps and then gently converges on both objectives, achieving high forgetting efficiency without triggering model collapse.

The three tasks pose different levels of challenges, as the curve slope in Figure \ref{fig:trade_off} varies with the volume of undesirable content. Copyrighted code unlearning is the most challenging from this point of view, because it requires LLMs to forget entire files. Remarkably, \ourapproach retains a notable advantage even on this most difficult task.

\paragraph{Application on Different LLMs}
We extend our evaluation to four widely-used code LLMs (\ie, CodeLlama-7B \cite{CodeLlama}, Qwen2.5-Coder-7B \cite{hui2024qwen2.5-coder}, Deepseek-coder-6.7B \cite{zhu2024deepseekcoder}, and Starcoder-7B \cite{starcoder}) to demonstrate the versatility of \ourapproach across diverse architectures and training recipes.
While holding all other experimental settings fixed, we track both forget quality and model utility over successive epochs on the copyrighted-code unlearning task.

\begin{figure}[t]
    \centering
    \includegraphics[width=0.42\textwidth]{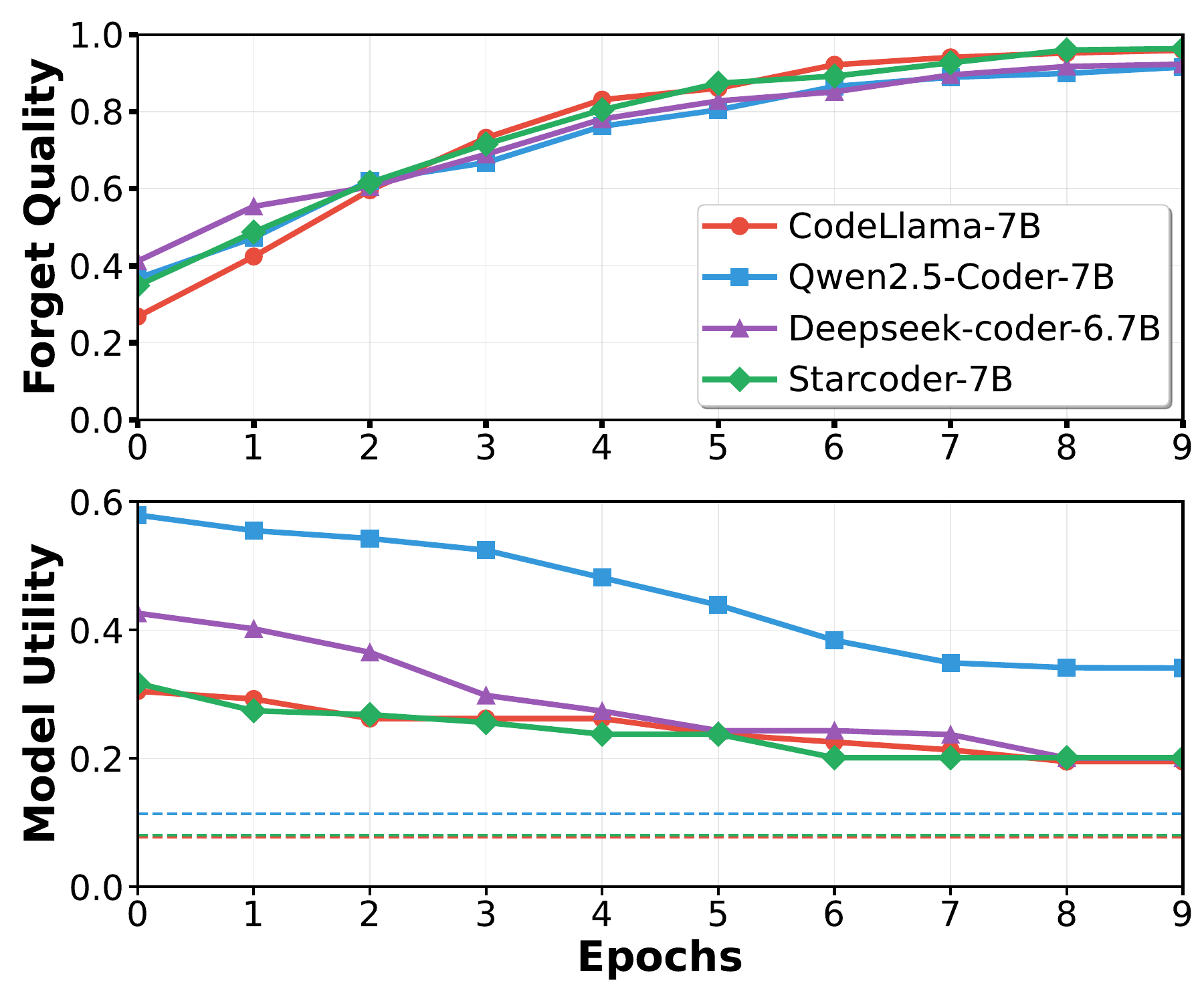}
    \caption{Performance of \ourapproach on different LLMs, where the dotted line represents the best model utility of baselines in achieving comparable (\ie, 90\%) forget quality.}
    \label{fig:LLMs}
\end{figure}

Figure \ref{fig:LLMs} reveals a consistent pattern across all four LLMs. As training proceeds, each LLM attains near-perfect forget quality while its utility curve stabilizes, which significantly outperforms all baseline approaches in model utility.
This uniformity underscores both the strength and robustness of \ourapproach. Among the evaluated models, CodeLlama-7B and StarCoder-7B exhibit the mildest utility degradation, implying that an LLM’s intrinsic stability can modulate the effectiveness of unlearning.

\subsubsection{Adversarial Attacks}
Recent studies have exposed the brittleness of unlearning: forgotten content can resurface when an adversary manipulates the prompt \cite{DBLP:journals/corr/abs-2408-10682,qi2025safety}.
We therefore assess \ourapproach against prefix injection attack, following previous work \cite{qi2025safety}.
In such an attack, an adversary prepends an increasing number of tokens from the copyrighted snippet to the prompt in an attempt to recover the erased snippets.
We conduct this evaluation on the copyrighted code task against CodeLlama-7B and measure robustness with the drop in forget quality.
To ensure validity of the experiment (\ie, the model remains practically useful), we only evaluate checkpoints that retain at least 60\% of the original LLM’s HumanEval performance.

\begin{figure}[t]
    \centering
    \includegraphics[width=0.45\textwidth]{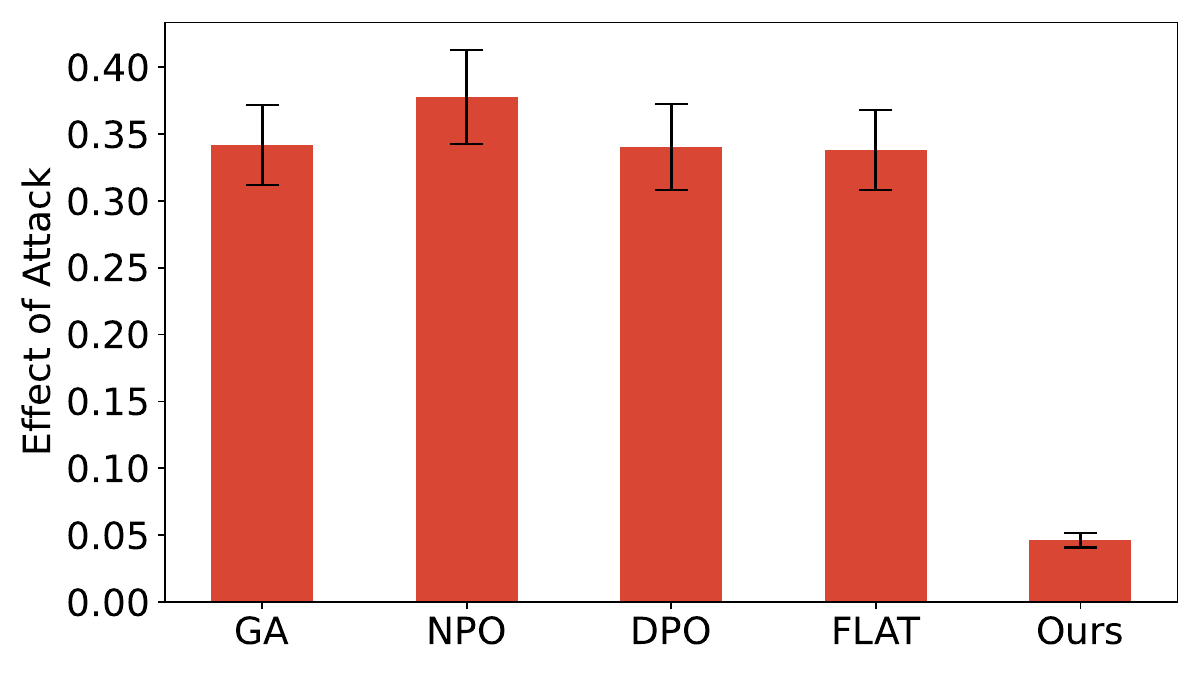}
    \caption{Comparison of adversarial attack results across different LLM unlearning approaches, showing mean attack effects with maximum and minimum ranges.}
    \label{fig:attack}
\end{figure}

Figure \ref{fig:attack} shows that \ourapproach is significantly more robust under prefix injection attack than other baselines.
Specifically, the similarity between the generation and the undesirable copyrighted code stays below 0.05, while others exceed 0.3. Moreover, \ourapproach shows the smallest variance among baselines, underscoring its stable resilience against prefix inject attack. These results provide clear evidence that \ourapproach delivers stronger and more reliable forgetting than baselines.

\begin{figure*}[ht!]
    \centering
    \begin{subfigure}[b]{0.32\textwidth}
        \centering
        \includegraphics[width=\textwidth]{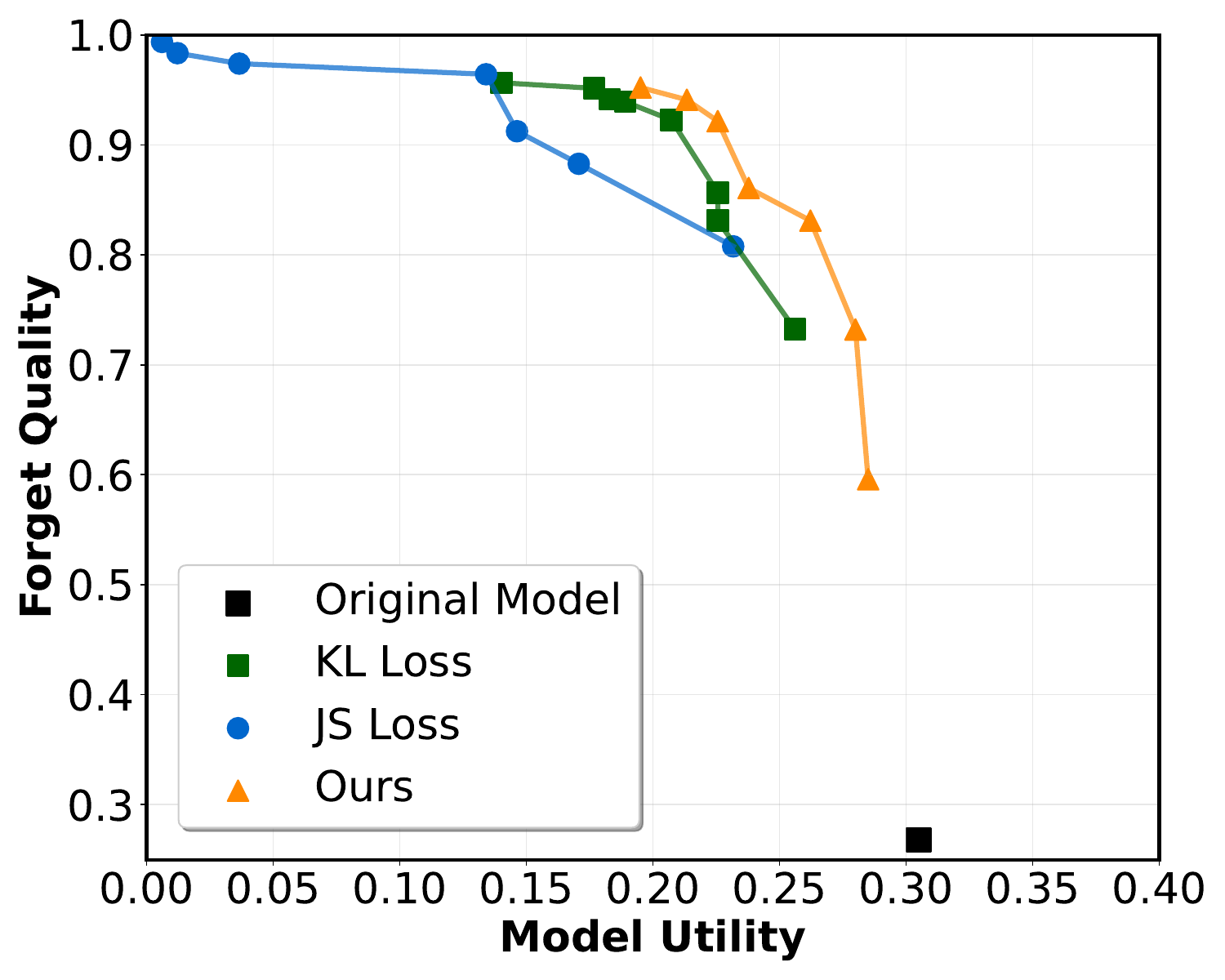}
        \caption{Loss Function}
        \label{KL_ablation}
    \end{subfigure}
    \hspace{0.005\textwidth}
    \begin{subfigure}[b]{0.32\textwidth}
        \centering
        \includegraphics[width=\textwidth]{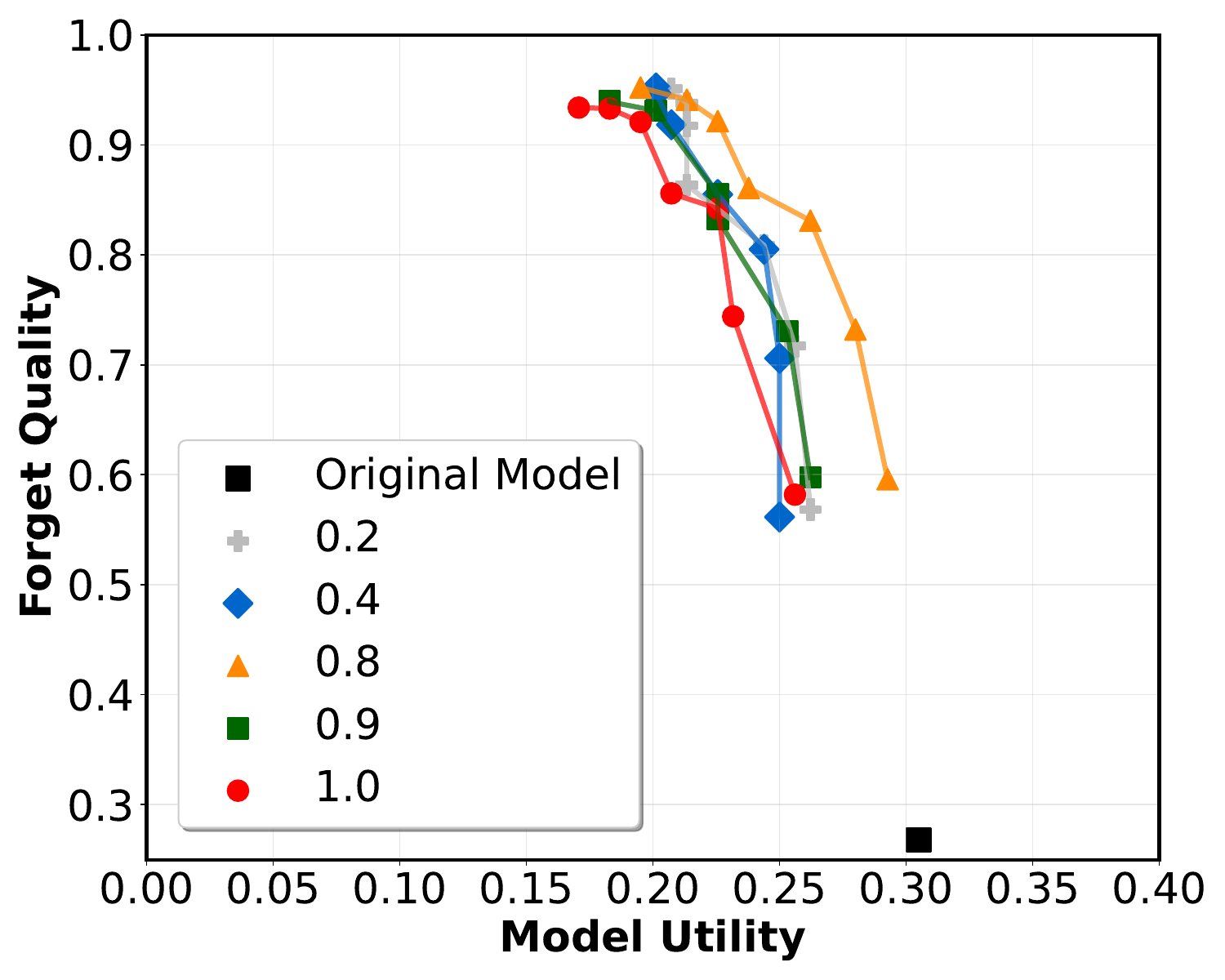}
        \caption{Hyperparameter $\mathbf{p}$}
        \label{fig:topp_ablation}
    \end{subfigure}
    \hspace{0.005\textwidth}
    \begin{subfigure}[b]{0.32\textwidth}
        \centering
        \includegraphics[width=\textwidth]{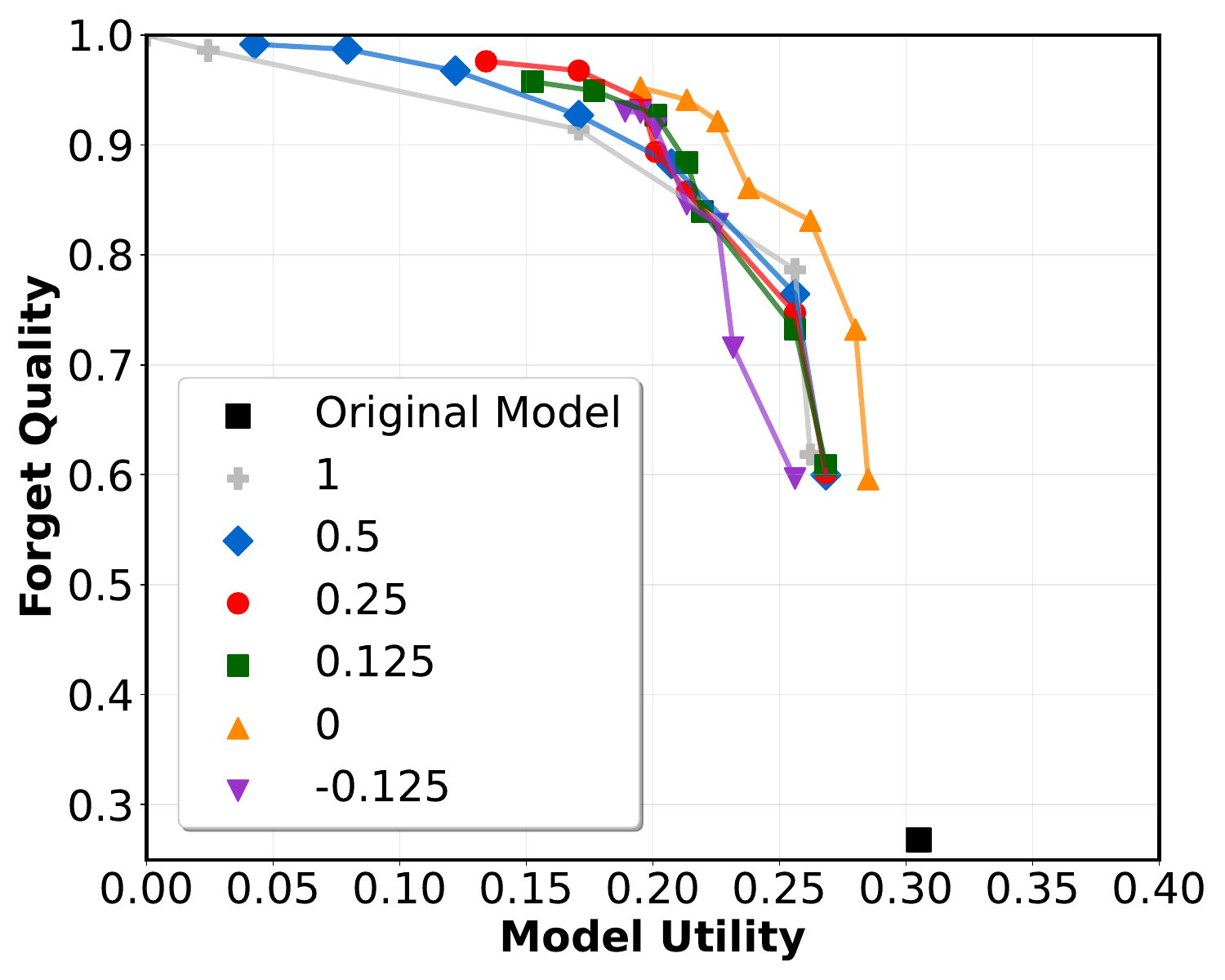}
        \caption{Hyperparameter $\mathbf{\alpha}$}
        \label{fig:hp_ablation}
    \end{subfigure}
    \caption{Ablation results on alternative loss function, and the impact of hyperparameters $p$ and $\alpha$ in \ourapproach.}
    \label{fig:ablation}
\end{figure*}

\paragraph{Ablation Study}
Finally, we perform some ablations.
We replace the default cross-entropy loss with two alternative divergence measures (\ie, KL and JS).
We also systematically vary the key hyperparameters $p$, which governs the strength of noise elimination, and $\alpha$, which scales the amplified suppression on forget data samples.

From Figure \ref{KL_ablation}, we observe that Cross-entropy outperforms both KL and JS divergence, delivering the best trade-off between forget quality and model utility.
Figure \ref{fig:topp_ablation} shows that the moderate noise elimination ($p=0.8$) achieves optimal results, while no noise elimination yields the worst.
Figure \ref{fig:hp_ablation} shows that $\alpha$ has a large impact on unlearning performance. When $\alpha=0$, which prohibits the generation of forget data by setting its probability to zero in the target distribution, the model achieves the most balanced performance. Positive $\alpha$ values are beneficial for deepening the degree of forgetting, while small negative $\alpha$ values delay forgetting to maintain usability.

\section{Conclusion}
In this paper, we have investigated existing LLM unlearning approaches on source code, identifying a significant utility degradation problem. To this end, we propose \ourapproach, a surgical, code-oriented unlearning method that forgets targeted code snippets while preserving other knowledge of programming languages intact. We also establish a benchmark covering copyrighted code, insecure code, and deprecated API unlearning tasks. Extensive experiments show \ourapproach significantly outperforms existing methods in both forgetting performance and user experience, while also exhibiting broad applicability and robustness against unlearning attacks.

\section{Acknowledgments}
This research is supported by the National Key R\&D Program under Grant No. 2023YFB4503801, the National Natural Science Foundation of China under Grant No. 62192733, 62192730, 62192731, and the Major Program (JD) of Hubei Province (No.2023BAA024).

\bibliography{aaai2026}

\appendix

\section{Preliminaries on LLM Unlearning Approaches}

\textbf{Gradient Ascent (GA) \cite{Unlearning}.} As a simple and widely used unlearning approach, GA \cite{Unlearning} performs gradient ascent on next-token prediction loss over the target forget data to approximately "reverse" the effects of gradient descent optimization that occurred during training on the forget data, formally denoted as $L_{GA}$. 

\begin{equation}
\begin{aligned}
    \mathcal{L}_{\text{GA}}(\Theta) &= -\underbrace{\mathbb{E}_{\mathcal{D}_f}[- \log(\pi_\Theta (y_f|x_f))]}_{\text{prediction loss}} \\
    &= \mathbb{E}_{\mathcal{D}_f}[\log(\pi_\Theta (y_f|x_f))].
\end{aligned}
\end{equation}

However, GA suffers from divergent properties and excessively rapid linear convergence rates, resulting in unstable training dynamics and potential model collapse. NPO is developed to address the limitation. NPO is inspired by DPO, which is a well-established approach in the field of LLM preference optimization.

\textbf{Direct Preference Optimization (DPO) \cite{DPO}.} DPO enables the model to learn to increase the probability of preferred responses $y_w$ while decreasing the probability of non-preferred responses $y_l$, all while optimizing under the constraint of staying close to a reference model. The loss function $\mathcal{L}_{\text{DPO}}(\Theta)$ is defined as:

\begin{equation}
\footnotesize{
\begin{aligned}
\mathcal{L}_{\text{DPO}}(\Theta)& = -\mathbb{E}_{(x,y_w,y_l)} \left[ \log \sigma \left( \beta \left( \log \frac{\pi_{\Theta}(y_w|x)}{\pi_{\text{ref}}(y_w|x)} \right. \right. \right. \\ & \qquad \qquad \qquad \left. \left. \left. - \log \frac{\pi_{\Theta}(y_l|x)}{\pi_{\text{ref}}(y_l|x)} \right) \right) \right],
\end{aligned}}
\end{equation}

where $(x,y_w,y_l)$ represents the training data triplets: prompt, preferred response, and non-preferred response, $\pi_{\Theta}$ is the model being trained, $\pi_{\text{ref}}$ is the reference model (typically the initial LLMs for training), $\beta$ is a hyperparameter that controls the degree of deviation from the reference model and $\sigma$ is the sigmoid function.

Although DPO is originally designed for LLM preference optimization, researchers have discovered that it can be applied to unlearning tasks by providing template responses $y_e$ (such as "I don't know") as the preferred response $y_w$, and $y_f$ as non-preferred response $y_l$.

\textbf{Negative Preference Optimization (NPO) \cite{NPO}.} Since unlearning tasks typically only provide negative samples (\ie, forget data), NPO retains only the log-probability ratio for non-preferred responses from the DPO formulation to steer the model away from generating undesired content. The NPO loss function $\mathcal{L}_{\text{NPO}}(\Theta)$ is defined as:

\begin{equation}
\mathcal{L}_{\text{NPO}}(\Theta) = -\mathbb{E}_{(x_f,y_f)} \left[ \log \sigma \left( \beta \left( - \log \frac{\pi_{\Theta}(y_f|x_f)}{\pi_{\text{ref}}(y_f|x_f)} \right) \right) \right],
\end{equation}
where NPO leverages the structural advantages of the DPO loss formulation (specifically the $\log \sigma$ transformation and the regularization provided by $\pi_{\text{ref}}$) to achieve logarithmic convergence rates, which typically yields more stable training dynamics compared to GA approach.

\textbf{Forget Data Only Loss Adjustment (FLAT) \cite{FLAT}.} FLAT proposes to use a loss function that balances two key objectives: discouraging the generation of undesirable responses to forget data $y_f$ while simultaneously promoting the production of appropriate template responses $y_e$ when encountering relevant forget prompts,

\begin{equation}
    \mathcal{L}_{\text{FLAT}}(\Theta)  = - \lambda_f \cdot L_f(x_f, y_f; \Theta) + \lambda_e \cdot L_e(x_f, y_e; \Theta),
\end{equation}
where $\lambda_f$ and $\lambda_e$ are wight to balance between $L_e(x_f, y_e; \Theta)$ and $L_f(x_f, y_f; \Theta)$. FLAT introduces f-divergence to establish an appropriate equilibrium between the two objectives. The loss function is subsequently reformulated into a variational form of f-divergence:

\begin{equation}
\small{
\begin{aligned}
     \mathcal{L}^{\text{FD}}_{\text{FLAT}}(\Theta) &= - \mathbb{E}_{(x_f,y_e,y_f)} \left[g^*(\mathbb{P}(x_f, y_e; \Theta)) + f^*(g^*(\mathbb{P}(x_f, y_f; \Theta)))\right], \\
    & \mathbb{P}(x_f, y; \Theta) = \frac{1}{|y|} \sum_{t=1}^{|y|} \frac{\pi_{\Theta}(y^{1:t} \mid x_f)}{\pi_{\Theta}(y^{1:t-1} \mid x_f)},
\end{aligned}}
\end{equation}
where \(y\) represents the template answer \(y_e\) or \(y_f\), and \(y^{1:t}\) denotes the first \(t\) tokens of the sequence \(y\). FLAT automatically assigns appropriate weights $g^*$ and $f^*(g^*())$ to the loss components. f-divergence is a divergence framework that includes various divergence measures, such as the Kullback-Leibler (KL) divergence, Jensen-Shannon (JS) divergence, and Pearson divergence.
$g^*$ and $f^*(g^*())$ can be configured according to the specific divergence measure chosen.

\section{Implementation Details} \label{sec:implement}
We detail the specific implementation details of our evaluation from three aspects. Specifically,

\textbf{Configurations for Downstream Tasks.}
For copyrighted code unlearning and insecure code unlearning tasks, following previous unlearning work \cite{Unlearning, NPO,FLAT}, we first train the model on forget data $D_{FG}$ to ensure the model has memorized the content before applying unlearning approaches. We use a learning rate of 2e-5, a batch size of 32, and train for 10 epochs. To prevent overfitting on $D_{FG}$ during training, we employ continued pretraining to train the model, where in each epoch, the forget data is mixed with different pretraining data in a ratio of approximately 1:9, which is a subset of The Stack corpus.
For deprecated API unlearning task, we directly apply LLM unlearning approaches on the LLMs without training on forget data. This is because the APIs we target in this task are common functions from widely used libraries that LLMs have basically encountered during pretraining.

\textbf{Training Setup for Unlearning.}
We use CodeLlama-7B \cite{CodeLlama} as our base model by default for our evaluation. All experiments are conducted on 4 NVIDIA A100 GPUs. Following previous work \cite{NPO}, we set the batch size to 32 and train for 10 epochs.
For all unlearning approaches, we select the learning rate via grid search from the set \{1e-4, 5e-5, 1e-5, 5e-6, 1e-6\}, optimizing for the forget quality metric on forget data. 
We use AdamW optimizer \cite{adamw} with a weight decay of 0.01 and set the maximum sequence length to 1024 tokens during training.
During training, we only optimize the loss over the forgetting part of each example.
The unlearning training is conducted five times using different random seeds, and the final evaluation results are averaged across the five
runs.

\textbf{Configurations for Baselines and \ourapproachbf.}
NPO and DPO use the hyperparameter $\beta $ of 0.1. FLAT employs KL divergence in its loss, which demonstrated superior performance in the original study, where $g^*(v)=v$ and $f^*(u)=e^{u-1}$. For our approach \ourapproach, we set $p$ and $\alpha$ to 0.8 and 0, respectively.

\section{Perceptual Quality Evaluation}

Following the work \cite{macketanz-etal-2022-perceptual,DBLP:conf/icmcs/ZhangLSLMZ23}, we conduct a perceptual quality evaluation to assess the subjective human experience of our approach. 

\textbf{Setup.}
We invite three independent volunteer evaluators (each with two years or more of software development experience). We randomly select 20 samples from each of the three downstream tasks, including generation results from different approaches. The evaluation is conducted in a comparative manner, where evaluators are presented with samples from our approach and baseline approaches, and asked to select the better one with a justification. Each human evaluator assesses all samples without knowing which approach produced each sample. We calculate the win rate of our approach against baselines.
The win rate is calculated as the total number of times our approach outperformed baselines divided by the total number of comparisons in the evaluation.

Given that human evaluation can only assess a limited portion of samples due to practical constraints, and recent research suggests that LLMs can substitute human evaluation to some extent with high evaluation consistency \cite{liu-etal-2023-g}, we introduced GPT-4 to evaluate all samples across all downstream tasks.

\begin{table}[htbp]
\centering
\caption{Win rate comparison on perceptual evaluation.}
\label{tab:win_rate}
\begin{tabular}{lcc}
\toprule
\textbf{Comparison} & \textbf{Human Evaluation} & \textbf{GPT-4 Evaluation} \\
\midrule
Ours vs GA & 81\% & 77\% \\
Ours vs DPO & 92\% & 86\% \\
Ours vs NPO & 76\% & 73\% \\
Ours vs FLAT & 87\% & 81\% \\
\bottomrule
\end{tabular} \label{tab:perceptual}
\end{table}

\textbf{Results.}
The results of the perceptual quality evaluation are shown in Figure \ref{tab:perceptual}. We find that our approach achieves win rates above 70\% compared to baselines. According to the reasons provided by human evaluators, the common feedback was that our approach still produced valuable outputs after forgetting specific content, while baselines often produced empty outputs or messy code.

\section{Extended Related Work}

\subsection{Code Generation with LLMs}
Code generation is the computational process of automatically producing executable source code from specifications or natural language inputs.
Generative models such as ChatGPT~\cite{achiam2023gpt4} and Claude~\cite{claude3.5} have demonstrated strong capabilities in code generation, benefiting from pretraining on large-scale data and code repositories.
Given the immense potential of LLMs, researchers have developed code-specialized LLMs, such as CodeX~\cite{codex}, CodeLlama~\cite{CodeLlama}, DeepSeek-Coder \cite{zhu2024deepseekcoder}, Qwen2.5 Coder~\cite{hui2024qwen2.5-coder}, and CodeGemma~\cite{team2024codegemma}, which further enhanced LLMs' capabilities across various software development tasks. 
However, several critical challenges remain unresolved in leveraging LLMs for code generation, including concerns about safety, legality, and reliability. 

Recently, an emerging concern is that LLM-generated code may inadvertently violate security best practices. Because LLMs learn from massive public code repositories, many of which include buggy or outdated patterns, they can propagate known vulnerabilities or poor coding habits. Researchers have begun exploring security-focused code generation techniques~\cite{safecoder, mohsin2024can, wu2024versicode, wang2024and, xu2024first, fu2023security}. Representative approaches include SaferCode~\cite{safecoder}, which augments LLM training with carefully curated secure code datasets, and work~\cite{mohsin2024can}, which proposes incorporating security examples directly into prompts to avoid unsafe coding patterns. Beyond security issues, VersiCode~\cite{wu2024versicode} introduces a version-specific benchmark to evaluate how models handle API changes across different releases, revealing that version-controllable code generation poses significant challenges. Failing to adapt can lead to functional errors and long-term maintenance issues. Similarly, SecuCoGen~\cite{wang2024and} systematically analyzes deprecated API usage in LLM-based code completion. Testing seven advanced models on over 28k prompts and 145 API changes, they observe a substantial rate of outdated suggestions, indicating that many LLMs fail to migrate to newer APIs. Additionally, copyright concerns have also been raised regarding LLM-generated code. LiCoEval~\cite{xu2024first} systematically evaluates LLMs’ license compliance and finds that models sometimes produce code fragments that closely resemble copyrighted implementations, often without proper attribution.

These limitations highlight that despite remarkable progress, LLM-based code generation still falls short of developers' expectations for producing secure, legally compliant, and up-to-date code. Our work aims to address these gaps by proposing an effective unlearning approach for suppressing undesired code output from LLMs.

\subsection{LLM Unlearning}

Machine unlearning, the process of removing specific data and its influence from a trained machine learning model, has garnered significant attention in recent years. Since \cite{DBLP:conf/sp/CaoY15} first introduced the concept of machine unlearning, laying the groundwork for subsequent research. Due to the widespread application and the relative simplicity of classification models, they become the focal point for most of the machine unlearning works \cite{DBLP:conf/cvpr/GolatkarAS20,DBLP:journals/corr/abs-2002-02730,DBLP:conf/aistats/IzzoSCZ21,DBLP:conf/sigmod/SchelterGD21,DBLP:conf/sp/BourtouleCCJTZL21,DBLP:conf/nips/SekhariAKS21}.
Machine unlearning approaches can be broadly categorized into three types: data-reversed training \cite{DBLP:journals/tnn/TarunCMK24,DBLP:conf/infocom/LiuFCLMWM22,DBLP:journals/tifs/ChundawatTMK23}, influence function based approaches \cite{DBLP:conf/aistats/IzzoSCZ21}, and optimization based unlearning \cite{DBLP:conf/icml/GuoGHM20,DBLP:conf/alt/Neel0S21}. Data-reversed training techniques employ noise injection to reverse or weaken the model's memory and dependence on specific data. Influence function-based approaches evaluate the impact of training samples on the model by perturbing model parameters, which are less common in LLM unlearning due to the computational complexity of Hessian matrix inversion. Optimization-based unlearning, which is the focus of our work, achieves forgetting through direct model optimization.
The rapid advancement of LLMs presents more significant challenges for machine unlearning. Their massive parameter scales and generative nature make traditional unlearning approaches unsuitable \cite{DBLP:journals/natmi/LiuYJCBHYLXLVBKL25}. Recent research efforts have been dedicated to the domain of LLM unlearning within the field of NLP, proposing methods such as GA \cite{Unlearning}, DPO \cite{DPO}, NPO \cite{NPO}, and FLAT \cite{FLAT}, which aim to address forgetting tasks in the field of NLP, such as user privacy unlearning, copyrighted books unlearning, and harmful content unlearning \cite{Unlearning}.
To facilitate the evaluation of LLM unlearning, several benchmarks have been proposed. These include synthetic profiles and question-answer pairs about fictitious authors \cite{DBLP:journals/corr/abs-2401-06121}, as well as the forgetting of specific content like Harry Potter books \cite{DBLP:journals/corr/abs-2310-10683}. Moreover, to assess the vulnerabilities of unlearned models, various attack approaches have been employed, such as prefix injections \cite{Jailbroken,DBLP:journals/corr/abs-2408-10682,qi2025safety}. These attacks help researchers understand the robustness of unlearning techniques and identify areas for improvement.

Given that source code possesses unique characteristics compared to natural language, although the exploration of LLM unlearning in NLP has proven its value, its application in code generation remains largely unexplored. This necessitates the development of LLM unlearning approaches applicable to code to address the challenges.

\end{document}